\documentclass[a4paper,amsmath,amssymb,floatfix,pra,footinbib,twocolumn, unsortedaddress,showpacs,preprintnumbers]{revtex4}

\usepackage[pdftex]{graphicx}
\usepackage{dcolumn}
\usepackage{bm}
\usepackage{dsfont}
\usepackage{color}

\usepackage{url}
\usepackage[colorlinks=true ,citecolor=blue]{hyperref}

\newcommand{\ttt}{$\mathcal{T}_3$~}

\newcommand{\ii}{\text{i}}
\newcommand{\ee}{\text{e}}

\newcommand{\ket}[1]{|#1\rangle}

\def\bs#1{\boldsymbol{#1}}
\def\txt#1{\textrm{#1}}

\newcommand{\comment}[1]{}

\begin{document}
%
%

\title{Topology-induced phase transitions in quantum spin-Hall lattices}
\author{D. Bercioux}
\email{dario.bercioux@frias.uni-freiburg.de}
\affiliation{Freiburg Institute for Advanced Studies, Albert-Ludwigs-Universit\"at, D-79104 Freiburg, Germany}
\affiliation{Physikalisches Institut, Albert-Ludwigs-Universit\"at, D-79104 Freiburg, Germany}
\author{N. Goldman}
\affiliation{Center for Nonlinear Phenomena and Complex Systems - Universit$\acute{e}$ Libre de Bruxelles (U.L.B.), Code Postal 231, Campus Plaine, B-1050 Brussels, Belgium}
\author{D.~F. Urban}
\affiliation{Physikalisches Institut, Albert-Ludwigs-Universit\"at, D-79104 Freiburg, Germany}

\date{\today}

\begin{abstract}
Physical phenomena driven by topological properties, such as the quantum Hall effect, have the appealing feature to be robust with respect to external perturbations. Lately, a new class of materials has emerged manifesting  their topological properties at room temperature and without the need of external magnetic fields. These topological insulators are band insulators with  large spin-orbit interactions and exhibit the quantum spin-Hall (QSH) effect. Here we investigate the transition between QSH and normal insulating phases under topological deformations of a two-dimensional lattice. We demonstrate that the QSH phase present in the honeycomb lattice looses its robustness as the occupancy of extra lattice sites is allowed. Furthermore, we propose a method for verifying  our predictions with  fermionic cold atoms in optical lattices. In this context, the spin-orbit interaction is engineered via an appropriate synthetic gauge field. 
\end{abstract}

\maketitle

%
%

\section{Introduction}
In the last three decades the physics community has been fascinated by topological states of quantum matter, initiated by the discovery of the quantum Hall effect (QHE) by von Klitzing in 1981~\cite{klitzing:1980}. 
Recently it has been discovered that a new class of materials, referred to as \emph{topological insulators},  show robust conducting edge states even in the absence of external magnetic fields~\cite{kane:2005,bernevig:2006:a,bernevig:2006:b}. These materials are  two- or three-dimensional band insulators with a large spin-orbit interaction (SOI) and exhibit the so-called quantum spin-Hall (QSH) effect ~\cite{koning:2007,roth:2009,xia:2009,zhang:2009}. Here, spin-orbit effects play a role similar to the external magnetic field in the QHE.  However, contrary to an external magnetic field, SOIs conserve time-reversal symmetry (TRS), implying that the edge states appear as Kramers doublets or equivalently as counter-propagating modes along the system boundary carrying opposite spin. In two-dimensional (2D) systems a Z$_2$ topological invariant $\nu$ specifies the robustness of these helical edge-states in the presence of disorder and interactions~\cite{Xu:2006}. While $\nu=1$ states a topologically protected (\emph{i.e.} stabilized) QSH phase, $\nu=0$ indicates that the edge states are unstable with respect to impurity scattering therefore reducing the system to a normal insulator (NI). 

%
%
\begin{figure}[!t]
	\centering
	\includegraphics[width=\columnwidth]{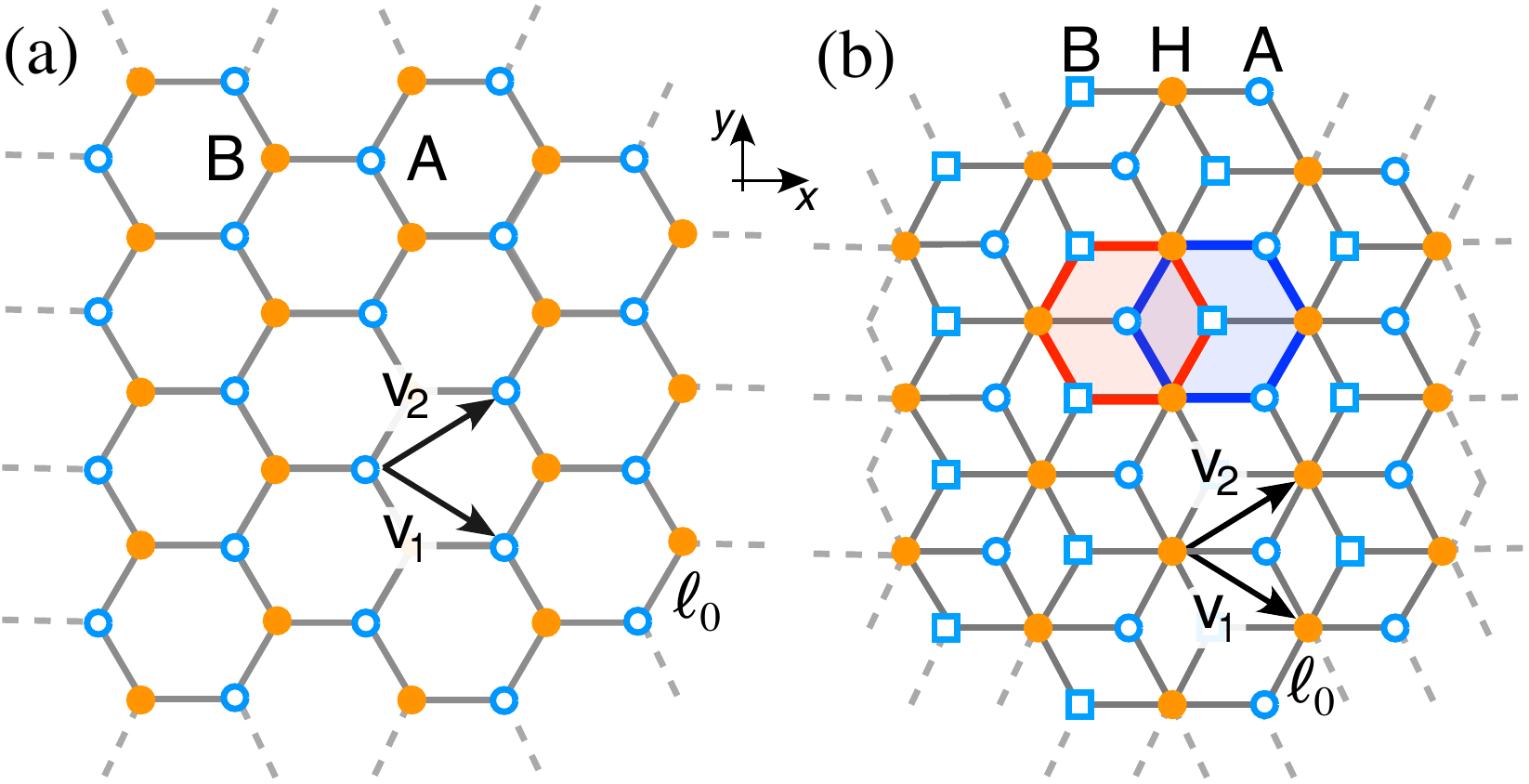}
	\caption{\label{fig:one}(Color online).  (a) HCL with lattice vectors $\mathbf{v}_1$ and $\mathbf{v}_2$ and two atoms in the unit cell. (b) \ttt lattice with three atoms in the unit cell. Rims A and B (open circle and square) have a lower connectivity of 3 compared to 6 for the hub site (filled circle).  The \ttt lattice can be viewed as two nested HCLs (emphasized by the red and blue hexagon).}
\end{figure}
%
%
In this Article we explore the interplay between spatial lattice topology, which is characterized by the lattice connectivity, and the topological order associated to the quantum spin-Hall phases. The latter  is characterized by the Z$_2$ topological invariant defined on a fibre bundle associated to the wave-functions over the first Brillouin zone~\cite{kane:2005}. We show that spatial deformations | modifying the lattice topology | induce a change in the Z$_2$ invariant, thus leading to a topological phase transition from a QSH insulator to a NI.

Specifically, we study the effect of adding an additional site to the unit cell of a honeycomb lattice (HCL) subjected to SOI. While this lattice deformation preserves the relativistic behavior of the low-energy modes, we show that the system undergoes a topological phase transition that changes its topological class, labeled by the Z$_2$ index, and its transport properties. The possibility to investigate this topological phase transition with ultra-cold atoms trapped in optical lattices is promising~\cite{Bloch:2008p894,Lewenstein:2007}. In this context, cold-atom experiments are not only suitable for engineering the QSH phase~\cite{Stanescu:2009,goldman:2010} and to test its stability with respect to controllable disorder~\cite{Sanchez:2010},  but also to induce lattice transformations.

%
%
\section{Model and formalism}
Adding an extra site to the unit cell of a HCL  gives rise to the so called \ttt lattice, c.f. Fig.~\ref{fig:one}.
The \ttt lattice not only differs from the HCL by the number of inequivalent lattice sites but also by their respective coordination numbers: while the HCL (Fig.~\ref{fig:one}a) is characterized by two inequivalent lattice sites A and B, each with coordination number 3, the \ttt lattice (Fig.~\ref{fig:one}b) has a unit cell with three inequivalent sites. Two of the latter (generally called \emph{rim} A and B) are characterized by a coordination number 3 while one site (referred to as \emph{hub} H) is connected to 6 nearest-neighbors~\cite{sutherland:1986,vidal:1998,korshunov:2001}.

Both lattices are modeled by a tight-binding (TB) Hamiltonian ${\cal H}_0=t \sum_{\langle n,m \rangle\alpha} c^\dag_{n\alpha} c_{m\alpha}$ with spin independent nearest-neighbor hopping amplitude $t$. Here, $c_{n\alpha}^\dag(c_{n\alpha})$ is the  creation (annihilation) operator for a particle with spin direction $\alpha$ on the lattice site $n$.
Regarding the SOI, symmetries allow for two types of interactions that preserve TRS,
namely the Rashba-type~\cite{kane:2005} and the Haldane-type~\cite{haldane:1988}. The effects of the Rashba-type SOI on the transport properties of the HCL and \ttt lattice have been reported in Refs.~\cite{kane:2005,bercioux:2004}. The Haldane-SOI, which is the interaction of interest in the present study, is modeled via a spin-dependent second-neighbor hopping term 
%
%
\begin{equation}\label{eq:SO}
\mathcal{H}_\text{SO} = \text{i}\,  t_\text{SO} \sum_{\alpha\beta}\!\!\sum_{\langle\langle n,m \rangle\rangle}
c_{n,\alpha}^\dag \left(\mathbf{d}_i\times\mathbf{d}_j\right)\cdot \bm{\sigma}_{\alpha\beta} \ c_{m,\beta}\, .
\end{equation}
%
%
Here, the $\bm{\sigma}_{\alpha\beta}$ are matrix elements of the Pauli-matrices $\bm{\sigma}$ with respect to the final and initial spin states $\alpha$ and $\beta$ and $\mathbf{d}_{i/j}$ are the two displacement vectors of the second-neighbor hopping process connecting sites $n$ and $m$.  Since in 2D lattices hopping is naturally restricted to in-plane processes, the SOI is effectively proportional to $\sigma_z$.

In the absence of SOI the energy spectra of the two lattices are characterized by two identical, electron-hole symmetric  branches~\cite{kane:2005,bercioux:2009}. Moreover, for the \ttt lattice a unique non-dispersive band is present at the charge neutrality point~\cite{bercioux:2009}. This is rooted in 
the lattice topology, which allows for insulating states with finite wavefunction amplitudes on the rim sites and vanishing amplitudes on the hubs.

%
%
\begin{figure}
	\begin{center}
	\includegraphics[width=\columnwidth]{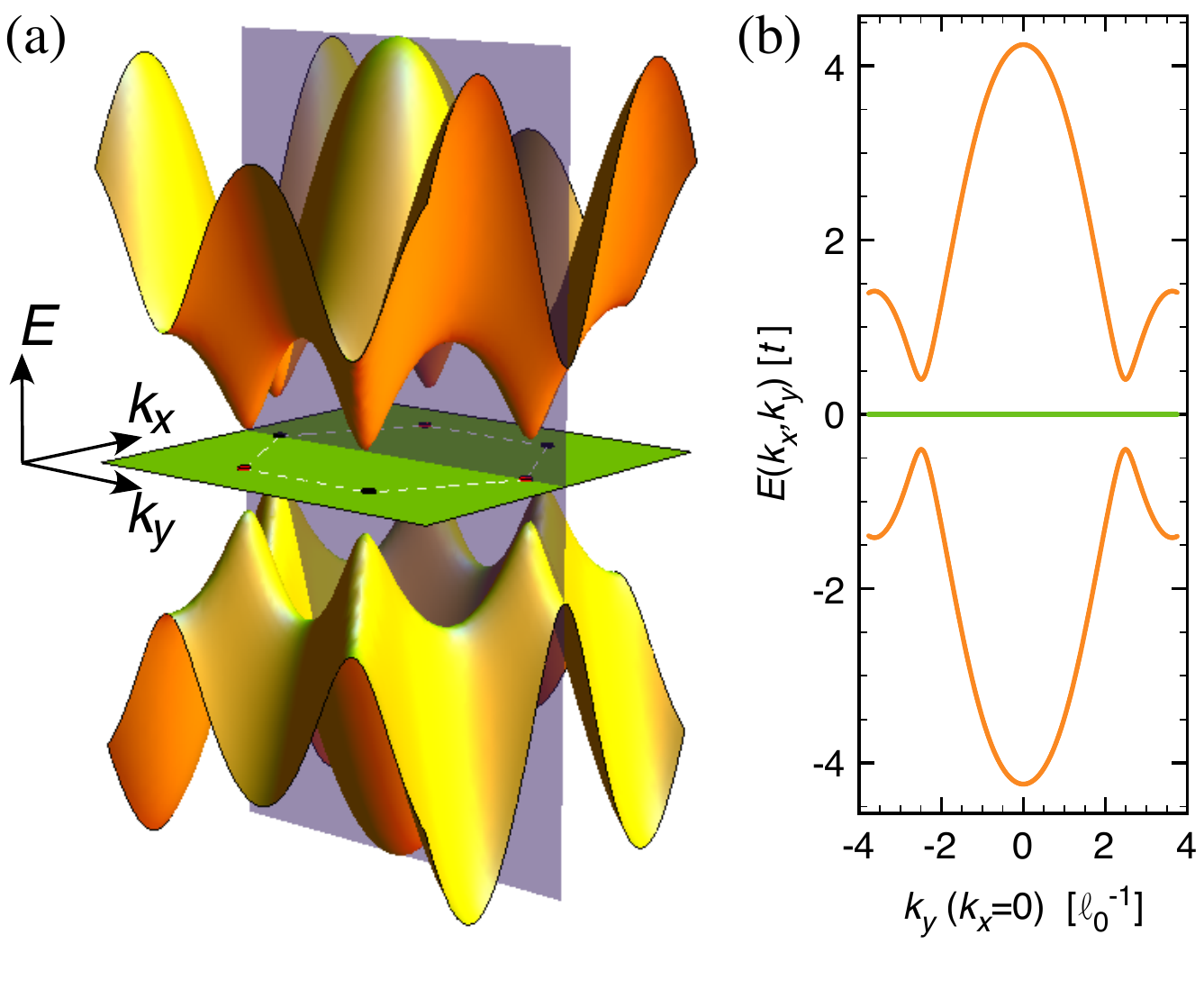}
	\caption{\label{fig:two} (Color online).  (a) Energy spectrum as a function of the in-plane momentum $(k_x,k_y)$ and for a fixed value of the spin-orbit interaction $t_\text{SO}=0.25t$. The first Brillouin zone and its six corners are indicated. (b) Cut of the energy spectrum at $k_x=0$ (shadowed plane in (a)).}
	\end{center}
\end{figure}
%
%

\section{Trivial \emph{vs.} topological insulators}
We start by analyzing the effect of finite SOI on the energy spectrum of the infinite \ttt lattice. 
Because of the unequal connectivity of rim and hub sites, the Hamiltonian~(\ref{eq:SO}) effectively induces hopping between  rim sites of the same kind while this is cancelled for hub sites.
The spectrum is obtained by exact diagonalization (cf.~App.~\ref{sec:two})  and the result is shown in Fig.~\ref{fig:two}. Due to its topological origin the non-dispersive band at the charge neutrality point is not affected by SOI. On the contrary, two bulk energy gaps open between the non-dispersive and the electron/hole branches, respectively. This property can be directly deduced from the single-valley long-wavelength approximation of the Hamiltonian (cf.~App.~\ref{sec:emd}). We find that the general form of this relativistic Hamiltonian | describing both the HCL and the \ttt lattice | reads
%
%
\begin{equation}\label{eq:DiracWeyl}
	{\cal H}=v_\text{F}{\bm\Sigma}\cdot\mathbf{p}\otimes \mathbb{I}_2 + \Delta_\text{SO}\Sigma_z\otimes\sigma_z\,.
\end{equation}
%
%
Here $v_\text{F}$ is the Fermi velocity, $\mathbf{p}=-\ii(\partial_x,\partial_y,0)$ the momentum operator ($\hbar=1$) in the lattice $xy$--plane, $\Delta_\text{SO}$ is the spin-orbit coupling constant, and $\mathbb{I}_2$ is the two-dimensional identity matrix. The pseudo-spin operator $\mathbf{\Sigma}$ for the HCL is given by the Pauli matrices $\mathbf{\Sigma}=(\sigma_x,\sigma_y,\sigma_z)$, thereby describing a $S=1/2$ pseudo-spin
system. On the contrary, for \ttt the pseudo-spin operator $\mathbf{\Sigma}=(\Sigma_x,\Sigma_y,\Sigma_z)$ is given by~\cite{bercioux:2009}
%
%
\begin{align}\label{eq:gm}
 \footnotesize\Sigma_x & =\frac{1}{\sqrt{2}}
 \begin{pmatrix}
 0 & 1 & 0 \\
 1 & 0 & 1 \\
 0 & 1 & 0
 \end{pmatrix},\
 \Sigma_y=\frac{1}{\sqrt{2}} \begin{pmatrix}
 0 & -\ii & 0 \\
 \ii & 0 & -\ii \\
 0 & \ii & 0
 \end{pmatrix},\ \\
\nonumber  & ~~~~~~~~~ \Sigma_z =\begin{pmatrix}
 1 & 0 & 0 \\
 0 & 0 & 0 \\
 0 & 0 & -1
 \end{pmatrix}. 
\end{align}
%
%
These matrices satisfy angular momentum commutation relations and describe an $S=1$ pseudo-spin (cf.~App.~\ref{sec:emd}).

\begin{figure}
	\centering
	\includegraphics[width=\columnwidth]{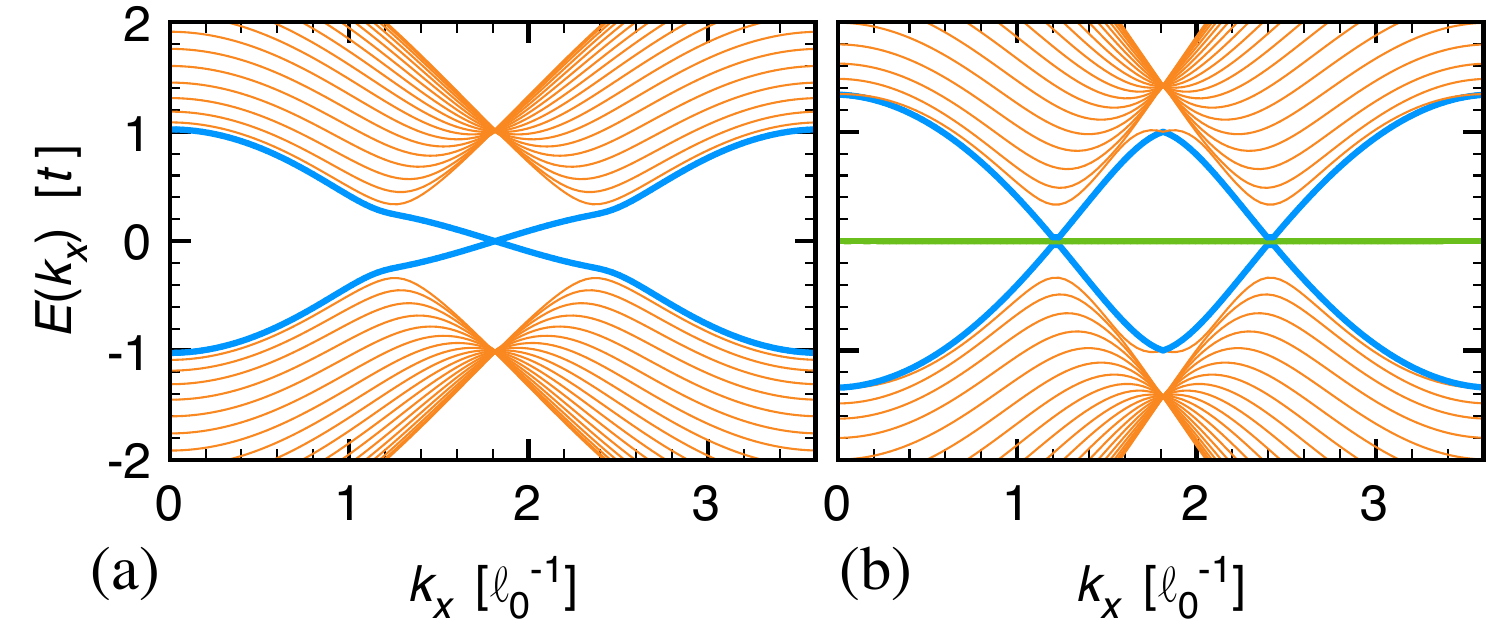}
	\caption{\label{fig:three} (Color online).  One-dimensional energy bands for a strip of (a) HCL and (b) \ttt lattice with $t_\text{SO}=0.05\,t$. The bands crossing the gap (blue lines) are spin filtered edge states.}
\end{figure}

A direct way of visualizing the SOI effects is to consider a finite piece of the lattice. The associated TB Hamiltonian can then be diagonalized with periodic boundary conditions imposed along one of the spatial directions. The HCL with finite SOI is characterized by the QSH phase:  for each energy value within the bulk energy-gap exists a single time-reversed pair of eigenstates on each edge of the lattice (see Fig.~\ref{fig:three}a) | these form two Kramers doublets, one on each edge. The conservation of TRS prevents the mixing of this couple of states due to small external perturbations and scattering from disorder~\cite{kane:2005}. 
The QSH phase is intrinsic to the bulk energy spectrum and can be validated by topological constants. Two important topological invariants are the $\text{Z}_2$ index $\nu$~\cite{kane:2005} and the spin Chern number $n_\sigma$~\cite{sheng:2006,fukui:2007}. In the absence of spin-mixing perturbations, both are related by $\nu=n_\sigma\text{mod}\,2$, where $n_\sigma=(N_{\uparrow}-N_{\downarrow})/2$ and $N_{\uparrow, \downarrow}$ represent the Chern numbers associated to the individual spins.
For the HCL it has been shown that $\nu=1$ therefore classifying it as a topological insulator. On the other hand it was found that SOI of Rashba-type does not induce this kind of edge-states~\cite{kane:2005}.

%
%
%
%
The physical picture is more involved in the case of a \ttt lattice with SOI. Here gapless edge states are also found for energies located within the gaps between the electron/hole band and the non-dispersive band, respectively (see Fig.\ \ref{fig:three}b).
However, contrary to the HCL case, we find two distinct crossing points for these edge states. For each energy value inside the bulk gaps two couples of time-reversed Kramers doublets are present, which therefore give rise to four pairs of states along the two edges. As a result, even if SOI allow for edge states within the bulk energy gap, these are not protected against disorder by TRS  | a scattering potential can couple a left-mover into a right-mover and \emph{vice versa}. This is confirmed by the topological invariants (cf.~App.~\ref{sec:z2}): we find that the $\text{Z}_2$-index vanishes and that the spin Chern number  $n_\sigma=2$.
For this reason the \ttt lattice is  topologically trivial.

A transition from the HCL to the \ttt lattice is accomplished by varying the hopping parameter  between hub and rim B from 0 to $t$. The remarkable result is that even for an infinitesimal coupling, the system is in the  trivial insulating phase. Therefore, we conclude that the different behavior of the HCL and the \ttt lattice with respect to the SOI arises from the different topologies of the two systems. The \ttt lattice can be viewed as two nested HCLs (see Fig.~\ref{fig:one}b). This supports the conjecture that two coupled topological insulators behave as a topologically trivial insulator~\cite{fukui:2007,fu:2007}. 
We can interpret the effect of lattice topology also from a different perspective. Although both lattices allow for a description in terms of a relativistic Hamiltonian~(\ref{eq:DiracWeyl}) in the long-wavelength approximation, only the HCL that is associated with a pseudo-spin 1/2 is a topological insulator whereas the \ttt lattice with pseudo-spin 1 is not. 
Nevertheless, the existence of other topological insulators associated with pseudo-spin 1~\cite{Franz:2010} suggests that the key factor  is not the dimension of the pseudo-spin but rather the \emph{total spin} of the systems under consideration, c.f. Tab.~\ref{tab:spin}.
In terms of the relativistic approximation~(\ref{eq:DiracWeyl}) the total spin is given by the combination of the valley-spin, accounting for the topology of the Bravais lattice,  the pseudo-spin, associated with the number of lattice sites in the unit cell, and the \emph{real} fermion spin. In the case of the HCL we have to combine three half-integer spins resulting in a half-integer total spin. On the contrary, for the \ttt lattice  we obtain an integer spin by combining two half-integer spins with an integer one. This conjecture is further supported by the fact that both the kagome lattice~\cite{guo:2009} and the Lieb lattice~\cite{Franz:2010} are non-trivial topological insulators, while the bilayer graphene is trivial~\cite{li:2010, fukui:2007}.

%
%
\begin{table}
\includegraphics[width=0.9\columnwidth]{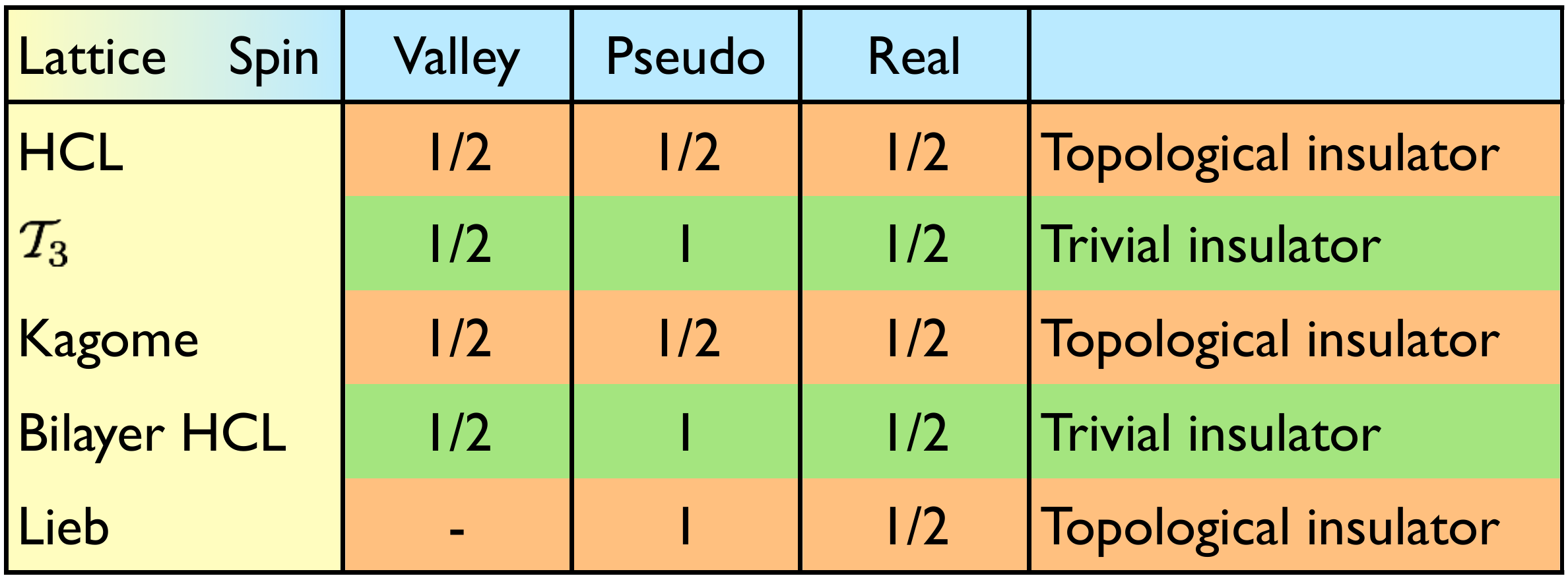}
\caption{\label{tab:spin} (Color online). In the table the valley-spin is associated with the topology of the Bravais lattice and the pseudo-spin with the number of lattice sites in the unit cell.}
\end{table}
%
%

\section{Optical lattice verification} 
Several proposals for realizing optical lattices resembling the HCL and the \ttt lattice have been put forward recently. The HCL lattice can be obtained by three coplanar plane waves with the same frequencies and a relative angle of 120$^\circ$~\cite{lee:2009}. On the other hand, the \ttt lattice can be realized considering three pairs of counter-propagating plane waves with a relative angle of 120$^\circ$ and a relative rotation of the polarization plane~\cite{rizzi:2006}.
Both set-ups share the same spatial configuration of the laser beams, therefore an adequate modulation of the laser intensity can  transform one lattice into the other.
The TB Hamiltonian  and their long wave length
approximations~(\ref{eq:DiracWeyl}) are valid for optical lattices
populated by  fermionic atoms, \emph{e.g.}
$^{40}$K or $^6$Li~\cite{Bloch:2008p894}. 

%
%
\begin{figure}[!t]
	\includegraphics[width=0.9\columnwidth]{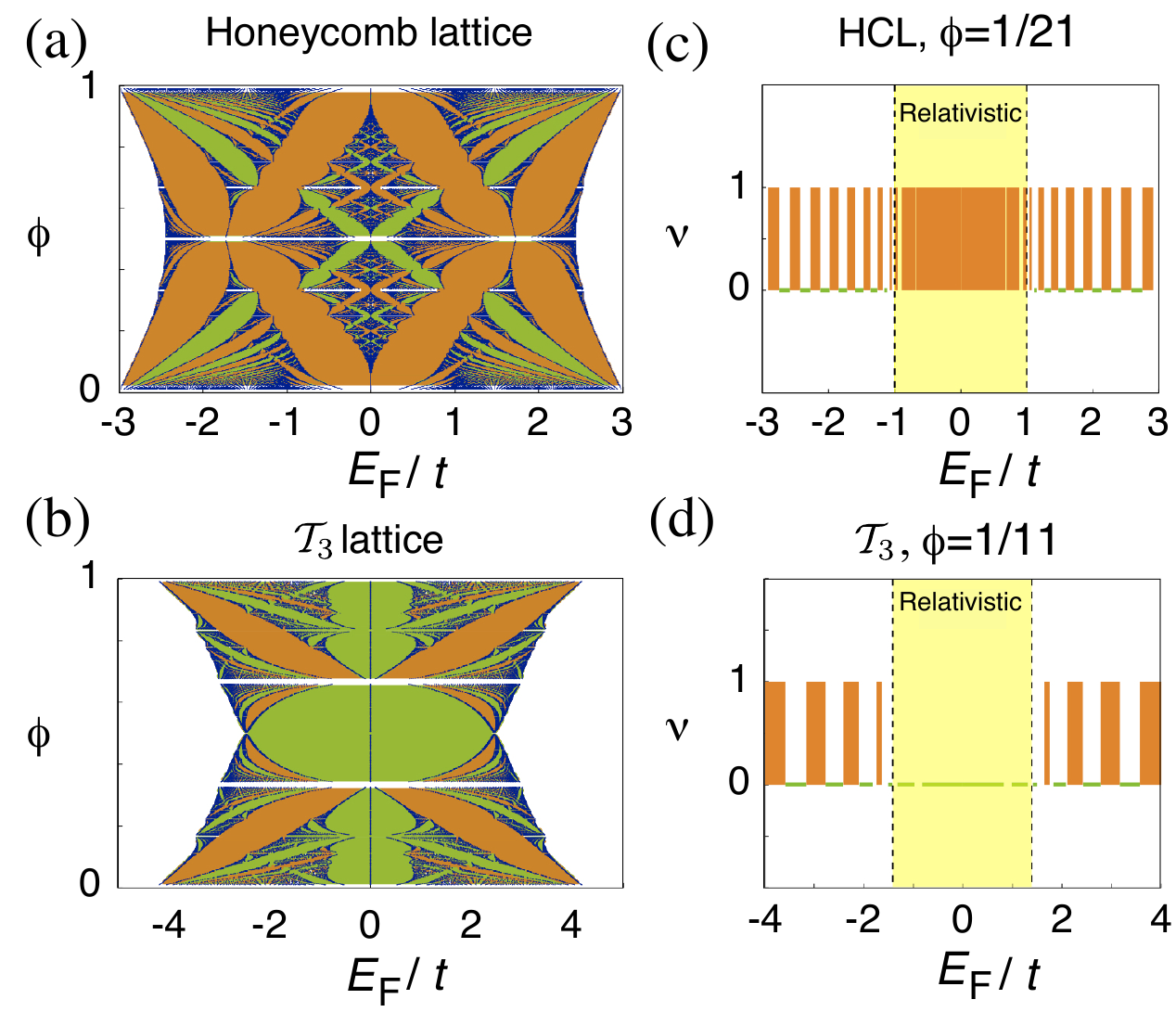}
	\caption{\label{fig:four} (Color online). (a) Phase diagrams in the Fermi energy-phase ($E_{\text{F}}-\phi$)--plane for the honeycomb lattice. The dark regions represent the energy spectrum. The gaps are colored according to the $\text{Z}_2$ index: orange [resp.\ green] gaps correspond to $\nu=1$ [resp.\ $\nu=0$].  (b) Phase diagrams in the ($E_{\text{F}}-\phi$)--plane for the \ttt lattice. (c) Cut of the phase diagram for the HCL in (a) for $\phi=1/21$. The orange bars denote the energy intervals for which $\nu=1$. 
(d) Cut of the phase diagram for the \ttt in (b) for $\phi=1/11$. 
Within the relativistic regime limited by $E_\text{VHS}=1$  [resp. $E_\text{VHS}\approx 1.4$], the gauge field~\eqref{eq:gauge} opens non-trivial [resp. trivial] Z$_2$ bulk gaps for the HCL [resp. $\mathcal{T}_3$ lattice].}
\end{figure}
%
%
Laser-assisted tunneling is a standard technique for implementing gauge fields for cold atoms, however a direct implementation of the Haldane-type SOI~(\ref{eq:SO}) is extremely complex. On the other hand, the QSH effect in 2D can be interpreted as two standard QHEs acting on each spin separately. Therefore, an alternative approach is to design an effective vector potential $\bm{A}=(A_x,A_y,A_z)\times\sigma_z$ that preserves TRS and opens a bulk energy gap.  Recently, this type of synthetic SU(2) field | practically realizable with state of the art techniques | has been proposed for a fermionic atom-chip forming a square lattice~\cite{goldman:2010}.
Here we propose to subject the atoms to a synthetic SU(2) gauge field 
%
%
\begin{equation}\label{eq:gauge}
\bm{A} = \left(0,\frac{2\pi \phi\, x}{\mathcal{S}},0\right)\times\sigma_z\,.
\end{equation}
%
%
that is associated with an opposite magnetic field for spin-up and spin-down electrons.
Here $\mathcal{S}$ is the area of the lattice plaquette and $\phi$ is the number of magnetic flux quanta per plaquette  felt by each spin, \emph{i.e.}, $\pm \phi$  for spin-up and spin-down, respectively (cf.~App.~\ref{sec:olp}). Note that the gauge field \eqref{eq:gauge} is space-dependent while preserving TR symmetry~\cite{bermudez:2010}.  The TB Hamiltonian including this gauge field for both the \ttt lattice and HCL in the long-wavelength approximation reads
%
%
\begin{equation}\label{eq:DiracWeyl:2}
	{\cal H}=v_\text{F}{\bm\Sigma}\cdot\mathbf{p}\otimes \mathbb{I}_2+\Delta (x) \Sigma_y \otimes \sigma_z\,.
\end{equation}
%
%
Contrary to the Hamiltonian~(\ref{eq:DiracWeyl}),  the SOI term contains $\Sigma_y$  and the coupling $\Delta (x)=2\pi \phi\, v_\text{F}\,\mathcal{S}^{-1} x$ is space-dependent. While the latter gives rise to a complex energy spectrum as a function of $\phi$,  we recover the features induced by the Haldane-type SOI in the ``low-flux" limit $\phi \ll 1$. 
The Z$_2$-phase diagrams for the HCL and \ttt lattice subject to the  synthetic gauge field~(\ref{eq:gauge}) are shown in Figs.~\ref{fig:four} (a) and (b), respectively. For arbitrary values of $\phi$, the gauge field leads to complex fractal structures featuring alternations of non-trivial (orange) and trivial (green) phases. However, we note that the HCL has a greater tendency to open QSH gaps than the \ttt lattice.
The relativistic regime | described by Eq.~\eqref{eq:DiracWeyl:2} |  is recovered in the low-flux limit $\phi \ll 1$~\cite{hatsugai:2006}. This regime is delimited by  the van Hove singularities (VHS)  $\pm E_{\text{VHS}}$ in the density of states. In Figs.~\ref{fig:four} (c) and (d), we show the Z$_2$ index $\nu$ as a function of the Fermi energy $E_\text{F}$ for $\phi \ll 1$. In the relativistic regime (yellow area), the HCL is characterized by QSH gaps with $\nu=1$, while the \ttt lattice is characterized by trivial insulating gaps with $\nu=0$. 
In this energy range we observe that the synthetic gauge field~\eqref{eq:gauge} reproduces the effects induced by the Haldane-SOI described above.
 Besides, we note that outside of the relativistic regimes, the HCL and \ttt lattice behave very similarly as they both feature alternation of non-trivial and trivial phases as a function of $E_\text{F}$. This important result emphasizes that the optical lattice topology dramatically influences the nature of the insulating phases within the relativistic regime.  Consequently, engineered gauge fields offer the unique possibility to explore Z$_2$-phase transitions as the optical-lattice topology is modified. 

Different schemes are available for distinguishing the QSH and normal insulating phases in a cold-atom experiment~\cite{goldman:2010}. In absence of spin-mixing perturbations the spin is conserved and the Z$_2$ topological constant reduces to $\nu=\vert N_{\uparrow} \vert\, \text{mod}\, 2$ because the individual spin-components are associated to $N_{\uparrow}=-N_{\downarrow}$. Therefore $\nu$ can be determined through density measurements and by applying the Streda formula~\cite{Umucalilar:2008}. Alternatively, the presence of helical edge-states can be also probed through light-scattering methods by exploiting the method reported in Ref.~\cite{Liu:2010}.

\section{Conclusions} We have studied the effect of Haldane-type SOI on two interrelated two-dimensional lattices: the honeycomb and the \ttt lattices. Both can be transformed  into each other by selectively switching on/off specific lattice sites. In absence of SOI, the main effect of the transition from HCL to \ttt is the appearance of a localized band at the charge neutrality point. For finite SOI though, a topological phase transition occurs that corresponds to the destruction of the QSH phase which is present in the HCL but not allowed in the \ttt\!\!. Since the occurrence of the QSH phase is related to the lattice topology, it is already destroyed by adding an infinitesimal weakly coupled lattice site to the HCL which gives rise to the \ttt lattice.
Finally, we have proposed a method for implementing this topological phase transition for fermionic cold atoms trapped in optical lattices. In this context, the QSH phases are obtained by synthesizing a gauge field reproducing the effects of the Haldane-type SOI for electrons.

\acknowledgments
We are in debt with I. Cirac, H. Grabert, M. Lewenstein, and M. Rizzi for useful discussions. DB is supported by the Excellence Initiative of the German
Federal and State Governments. NG thanks the F.R.S-F.N.R.S (Belgium) for financial support.

\appendix

\section{The bulk energy spectrum}\label{sec:two}
The energy spectrum of the \ttt lattice with Haldane-type~\cite{haldane:1988} SOI is obtained by diagonalization of the tight-binding Hamiltonian $\mathcal{H} = \mathcal{H}_0 + \mathcal{H}_\text{SO}$.
Here the first term is a nearest-neighbor hopping tight-binding Hamiltonian and includes the spin independent nearest-neighbor hopping amplitude $t$. The second term | defined in Eq.~\eqref{eq:SO} | describes the Haldane-type SOI.
Because of translational invariance,  this Hamiltonian can be diagonalized in reciprocal space.  The evaluation of the contribution due to the Haldane-SOI requires some care because of the different connectivity of hubs and rims. A second-next neighbor hopping process always interconnects sites of the same species (either rim A, rim B or hub). Because of the two possible distinct processes that connect two given hubs, the contributions arising from these two paths cancel. Therefore, the Haldane-SOI effectively only couples the rims of each species but not the hubs.

The Hamiltonian now can be written in matrix form
%
%
\begin{align}\label{eq:eight}
\mathcal{H}[\bm{k}] = 
t\begin{pmatrix}
0 &  \mathcal{A}^*  & 0  \\
\mathcal{A} & 0 & \mathcal{A}\\
0  &  \mathcal{A}^*  & 0
\end{pmatrix}  \otimes \mathbb{I}_2  
+ t_\text{SO}\begin{pmatrix}
\mathcal{B} & 0 & 0 \\
0& 0 & 0 \\
0& 0 & -\mathcal{B}
\end{pmatrix}
\otimes \sigma_z\,
\end{align}
%
%
where we have defined the coefficients 
%
%
\begin{align*}
\mathcal{A}&=1+ \ee^{\ii \bm{k}\cdot\mathbf{v}_1}+\ee^{\ii \bm{k}\cdot\mathbf{v}_2},\\
\mathcal{B}&=2\{ \sin(\bm{k}\cdot\mathbf{v}_1) - \sin(\bm{k}\cdot\mathbf{v}_2) + \sin[\bm{k}\cdot(\mathbf{v}_1+\mathbf{v}_2)]\}\,.
\end{align*}
%
%
Here  the translational vectors of the \ttt lattice are defined as $\mathbf{v}_1= (3/2;-\sqrt{3}/2)\ell_0$ and $\mathbf{v}_2= (3/2;\sqrt{3}/2)\ell_0$.
Equation~\eqref{eq:eight} corresponds to a six component spinorial wave function $\Psi_{\bm{k}}(\bm{r})=(\psi_\text{A},\psi_\text{H},\psi_\text{B})\otimes\sigma$, where we have introduced the components of the wave function on the two rims and the hub sites and the spin contribution $\sigma$.
Diagonalization of $\mathcal{H}[\bm{k}]$ yields the energy spectrum for the \ttt lattice in presence of Haldane-SOI
%
%
\begin{subequations}\label{eq:nine}
\begin{align}
\varepsilon_0(\bm{k})  = &0 \\
\varepsilon_\pm(\bm{k})= &\pm  \Big\{ 2 \big[ 3 (t^2 + t_\text{SO}^2) +
  2 t^2 (\cos[\bm{k}\cdot\mathbf{v}_1] \\ 
  & + \cos[\bm{k}\cdot(\mathbf{v}_1 -\mathbf{v}_2)] +
     \cos[\bm{k}\cdot\mathbf{v}_2])  \notag \\
     & - t_\text{SO}^2 (\cos[2 \bm{k}\cdot\mathbf{v}_1] -
     2 \cos[\bm{k}\cdot\mathbf{v}_2] \notag \\
     & +\cos[2 \bm{k}\cdot\mathbf{v}_2] +
     \cos[2 \bm{k}\cdot(\mathbf{v}_1-\mathbf{v}_2)] \nonumber \\
     &  +
     2 \cos[2\bm{k}\cdot\mathbf{v}_1 - \bm{k}\cdot \mathbf{v}_2]  +
     4 (\sin[\bm{k}\cdot\mathbf{v}_1] \notag\\ 
     &+ \sin[\bm{k}\cdot(\mathbf{v}_1-\mathbf{v}_2)] \sin[\bm{k}\cdot\mathbf{v}_2])) \big]\Big\}^{{1}/{2}}\nonumber
\end{align}
\end{subequations}
%
%
where $\bm{k}$ is the two-dimensional momentum vector and each band is two-fold degenerate.

\section{Long-wavelength approximation}\label{sec:emd}

For momenta close to the two independent $K$-points | $\mathbf{K} = 2\pi/3\ell_0 (1,-\sqrt{3}/3)$ and $\mathbf{K}' = 2\pi/3\ell_0 (1,+\sqrt{3}/3)$ | we can apply a long-wavelength approximation to the Schr\"odinger equation underlying the Hamiltonian $\mathcal{H}_0+\mathcal{H}_\text{SO}$.
This approximation consists in expressing the spatial part of the wave function as the product of a fast-varying part times a slow-varying part. In absence of perturbations that induce a mixing of the two $K$ points, we consider a single $K$--point | namely $\mathbf{K}$ |  and the wave function can be written as
%
%
\begin{equation}\label{eq:wf:lwa}
\Psi_\alpha(\bm{R}_\alpha) \propto \ee^{\ii \bm{k}\cdot\bm{R}_\alpha} F_\alpha^{\mathbf{K}}(\bm{R}_\alpha)\,.
\end{equation}
%
%
Here $\alpha\in\{\text{A,B,H}\}$ and $\bm{R}_\alpha$ is the lattice site coordinate. We substitute this wave-function into the Schr\"odinger equation and use the expansion 
%
%
\begin{equation*}\label{lwa}
F_\alpha^\mathbf{K}(\bm{R}_{\alpha'}\pm\bm{d}_j)\simeq F_\alpha^\mathbf{K}(\bm{R}_{\alpha'}) \pm \bm{d}_j \cdot \bm{\nabla}_{\bm{r}} \left.F_\alpha^{\mathbf{K}}(\bm{r})\right|_{\bm{r}=\bm{R}_{\alpha'}}\!\!\!\! + \mathcal{O}(|\mathbf{d}|^2)\,.
\end{equation*}
%
%
This approximation is valid for energies close to the charge neutrality point where the system energy is almost zero. In this case the associated Fermi wavelength $\lambda_\text{F}\sim 1/|\epsilon|$ is bigger than the modulus of the displacement vectors $\lambda_\text{F}\gg |\mathbf{d}_j|$. 
Collecting all the terms we are left with the  expression~\eqref{eq:DiracWeyl} where $v_\text{F}=3\ell_0t/2$ and $\Delta_\text{SO}= \sqrt{3} t_\text{SO}$. 

The pseudo-spin matrices $\bm{\Sigma}$ have been defined in \eqref{eq:gm}.
These matrices fulfill the algebra of the angular momentum $[\Sigma_i,\Sigma_j]=\ii \epsilon_{ijk}\Sigma_k$
and form a 3-dimensional representation of SU(2). However, contrary to the Pauli matrices, they do not form a Clifford algebra, \emph{i.e.}, \ $\{\Sigma_i,\Sigma_j\}\neq 2 \delta_{i,j} \mathbb{I}_3$.
By introducing a rotation operator around the $z$ axis defined by $\mathcal{D}_z(\phi) = \exp \left( -\ii \Sigma_z \phi \right)$, a generic state $\ket{\alpha}$ is transformed into itself by $\mathcal{D}_z(2\pi)\ket{\alpha}\to\ket{\alpha}$, implying that the pseudo-spin $\bm{\Sigma}$ describes an integer spin $S=1$.

\section{Topological properties of the \ttt lattice with Haldane-SOI}\label{sec:z2}

The Z$_2$ topological invariant characterizes the change in the Kramers degeneracy of a time-reversal symmetric system~\cite{kane:2005}. It can be deduced from the topological structure of the Bloch wave functions of the bulk crystal in the first Brillouin zone. Here, there are special points $\bm{k}=\text{M}_i$  that are time-reversal invariant and satisfy $-\text{M}_i=\text{M}_i+\mathbf{G}$ for a reciprocal-lattice vector $\mathbf{G}$. For the \ttt lattice  we have four of these points. These are defined by $\text{M}_{i=(n_1,n_2)}=1/2(n_1 \mathbf{g}_1 + n_2 \mathbf{g}_2)$ with $n_1,n_2\in\{0,\pm1\}$, where we have introduced the reciprocal lattice vectors $\mathbf{g}_1=2\pi/\ell_0(1/3;-\sqrt{3}/3)$ and $\mathbf{g}_2=2\pi/\ell_0(1/3;+\sqrt{3}/3)$.
Defining the time-reversal symmetry operator as $\mathcal{T}=\exp(\ii \pi \sigma_y/2)\mathcal{K}$ with the operator of complex conjugation $\mathcal{K}$, the four M points are found to fulfill \mbox{$\mathcal{H}(\text{M}_i)=\mathcal{T}\mathcal{H}(\text{M}_i)\mathcal{T}^{-1}$}.

In the case of the \ttt lattice the system is invariant under parity operation, which corresponds to an exchange of rim A with rim B and is expressed by the operator
%
%
\begin{equation}\label{parity}
\mathcal{P}= \begin{pmatrix}
0 & 0 & 1\\
0 & 1 & 0 \\
1 & 0 & 0
\end{pmatrix}\,.
\end{equation}
%
%
This implies that $[\mathcal{H}(\text{M}_i),\mathcal{P}]=0$ for all four time-reversal invariant momenta. According to Fu and Kane~\cite{fu:2007}  the Z$_2$ topological invariant $\nu$ is related to the parity eigenvalues $\xi_{2m}(\text{M}_i)$ of the $2m$-th occupied energy bands at the four time-reversal invariant points M$_i$ through the relation
%
%
\begin{equation}\label{z2}
(-1)^{\nu}= \prod_i \prod_{m=1}^N \xi_{2m}(\text{M}_i)\,.
\end{equation}
%
%
We have evaluated the eigenstates of $\mathcal{H}$ in the points $\text{M}_i$ analytically and therefore determined straightforwardly  the parity eigenvalues of the occupied bands. At filling $1/3$ and $2/3$ we find that all four eigenvalues are positive determining therefore a trivial Z$_2$ number $\nu=0$.

\section{Subjecting the Honeycomb and \ttt lattices to an SU(2) gauge field}\label{sec:olp}

On a lattice, the gauge field $\bs{A}$ enters the formalism through the Peierls (or Berry's) phases: as a particle hops from a site $i$ to a nearest-neighboring site $j$, the wave function acquires a non-trivial phase $\ee^{\ii \theta_{i j}}$, where $\theta_{ij}=\int_i^j \bs{A} \cdot \bs{\txt{d}l}$.
The set of Peierls phases $\{ \theta_{ij} \}$, defined on the whole lattice, therefore witnesses the presence of a specific gauge field $\bs{A}$. The physical observables associated to the gauge field, or equivalently to the set $\{ \theta_{ij} \}$, are given by the loop operators defined on each plaquette $\square$:
\begin{equation}
W(\square)=\ee^{\ii \theta_{1 2}} \ee^{\ii \theta_{2 3}} \dots \ee^{\ii \theta_{N 1}} ,
\end{equation}
where the plaquette $\square$ is delimited by the sites $\{ 1, 2, \dots , N \}$. In the case of an Abelian or U(1) gauge field, the loop operator is related to the magnetic flux $\Phi (\square)$ penetrating the plaquette through the relation $W(\square)=\ee^{\ii \Phi (\square)}$. 

In this work, we considered a gauge field of the following form
%
%
\begin{equation}
\bs{A}= \frac{1}{\mathcal{S}} \bigl ( 0 , 2 \pi \phi x , 0 \bigr )\otimes \sigma_z = \begin{pmatrix}  \bs{A}_{\uparrow} & 0 \\ 0  & \bs{A}_{\downarrow} 
\end{pmatrix}, \label{mygauge}
\end{equation}
%
%
that does not couple the two spin species together. Hence up and down spins feel individual and opposite magnetic fluxes
%
%
\begin{align}
\Phi_{\uparrow}(\square)&=\int_{\square} \bs{\nabla}\times \bs{A}_{\uparrow}=-\int_{\square} \bs{\nabla}\times \bs{A}_{\downarrow}=-\Phi_{\downarrow}(\square) \\ \notag
&=2 \pi \phi ,
\end{align}
%
%
that are uniform. The loop operators corresponding to the gauge field \eqref{mygauge} thus yield $W=\ee^{\ii 2 \pi \phi \sigma_z}$ within every plaquette of the lattice. The helical edge-states, hallmark signature of the QSH phase, are a direct consequence of the double-Hall system produced by the specific gauge field \eqref{mygauge} which preserves time-reversal symmetry. 

%
%
\begin{figure}[!t]
	\begin{center}
	\includegraphics[width=\columnwidth]{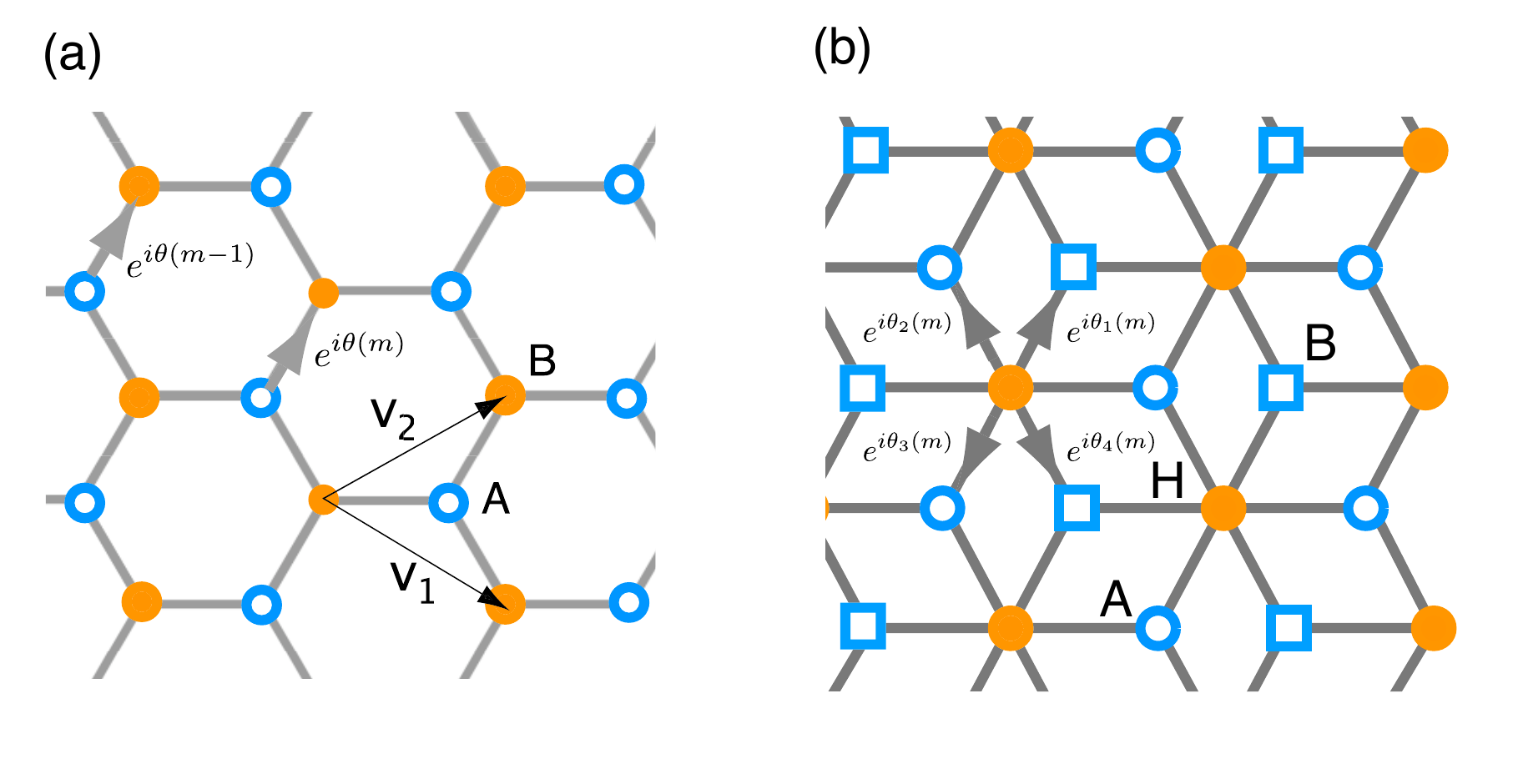}
	\caption{\label{NGfig} (Color online). Peierls phases generating the SU(2) gauge field for (a) the honeycomb lattice and (b) in the \ttt lattice.}
	\end{center}
\end{figure}
%
%

The gauge field \eqref{mygauge} can be engineered in an optical-lattice experiment by inducing the appropriate set of Peierls phases $\{ \theta_{ij} \}$ such that
%
%
\begin{equation}
W=\ee^{\ii \theta_{1 2}} \ee^{\ii \theta_{2 3}} \dots \ee^{\ii \theta_{N 1}}=\ee^{\ii 2 \pi \phi \sigma_z}   ,\label{myloop}
\end{equation}
%
%
for all the plaquettes constituting the honeycomb and \ttt lattices. Note that the Peierls phases can be generated through different schemes, based on the Raman transitions produced by additional lasers or fields along the appropriate links~\cite{jaksch:2003,Osterloh:2005,goldman:2010}. An adequate set of Peierls phases $\{ \theta_{ij} \}$ satisfying Eq.~\eqref{myloop}, and thus reproducing the effects of the gauge field \eqref{mygauge} on the honeycomb and \ttt lattices, are represented in Fig.~\ref{NGfig}.

In the case of the HCL, a single space-dependent Peierls phase $\theta (m)=2 \pi \phi m \sigma_z$ is necessary, as illustrated in Fig.~\ref{NGfig}a. The single-particle Schr\" odinger equation then takes the form of two coupled Harper difference equations for the wave-function $\Psi= \left(\psi_\text{A},\psi_\text{B}\right)$ defined at the site $(m,n)$,
%
%
\begin{subequations}\label{harper1}
\begin{align}
\psi_\text{A} (m,n)= & \psi_\text{B} (m,n-1) + \psi_\text{B} (m+1,n-1)\\
 & + \ee^{\ii 2 \pi \phi m \sigma_z} \psi_\text{B} (m,n) , \notag \\
\psi_\text{B} (m,n)= & \psi_\text{A} (m,n+1) + \psi_\text{A} (m-1,n+1) \\
& + \ee^{-\ii 2 \pi \phi m \sigma_z} \psi_\text{A} (m,n) \notag, 
\end{align}
\end{subequations}
%
%
and the loop operator indeed equals \comment{$W=\ee^{\ii 2 \pi \phi m \sigma_z} \ee^{-\ii 2 \pi \phi (m-1) \sigma_z}=\ee^{\ii 2 \pi \phi \sigma_z}$} $W=\ee^{\ii 2 \pi \phi \sigma_z}$, within all the hexagonal plaquettes. \\

In the case of the \ttt lattice, four space-dependent Peierls phases are required,
%
%
\begin{subequations}
\begin{align}
\theta_1 (m)&= 2 \pi \phi (3 m - 1/2) \sigma_z, \\ 
\theta_2 (m)&= 2 \pi \phi (3 m - 1) \sigma_z ,  \\
\theta_3 (m)&= 2 \pi \phi (- 3 m - 1/2) \sigma_z,\\ 
\theta_4 (m) &= 2 \pi \phi (- 3 m - 1) \sigma_z ,
\end{align}
\end{subequations}
%
%
as represented in Fig.~\ref{NGfig}b. The Harper equations for the wave-function $\Psi= \left( \psi_\text{A}, \psi_\text{B}, \psi_\text{H}\right)$ defined at the site $(m,n)$ yields
%
%
\begin{subequations} \label{harper2}
\begin{align}
\psi_\text{A} (m,n) = & \ee^{-\ii \theta_3 (m)} \psi_\text{H} (m,n) + \psi_\text{H} (m,n-1)\\ 
& + \ee^{-\ii \theta_2 (m+1)} \psi_\text{H} (m+1,n-1), \notag \\
\psi_\text{B} (m,n) = & \ee^{-\ii \theta_1 (m)} \psi_\text{H} (m,n) + \psi_\text{H} (m,n+1) \\
& + \ee^{-\ii \theta_4 (m-1)} \psi_\text{H} (m-1,n+1), \notag \\
\psi_\text{H} (m,n) =& \ee^{\ii \theta_3 (m)} \psi_\text{A} (m,n) + \psi_\text{B} (m,n-1) \\
&  +\ee^{\ii \theta_1 (m)} \psi_\text{B} (m,n) + \psi_\text{A} (m,n+1)\notag\\
& + \ee^{\ii \theta_4 (m)} \psi_\text{B} (m+1,n-1)  \notag \\
& + \ee^{\ii \theta_2 (m)} \psi_\text{A} (m-1,n+1) \notag
\end{align}
\end{subequations}
%
%
and one verifies that $W(\square)=\ee^{\ii 2 \pi \phi \sigma_z}$ within each rhombus forming the \ttt lattice. 

The energy spectra of the honeycomb and \ttt lattices are readily obtained by diagonalizing the Harper equations \eqref{harper1} and \eqref{harper2} numerically, leading to the butterfly-like structures represented in Figs.~\ref{fig:four}a and b of the manuscript. Since the gauge field does not couple the two spin components, the Z$_2$-topological invariant $\nu= \vert N_{\uparrow} \vert \text{mod}\, 2$ is directly obtained by computing the Chern number $N_{\uparrow}$ associated to the up-spin. This Chern number can be evaluated numerically by considering one copy of the two-fold Harper equations~\cite{hatsugai:2006}. Considering the more general relation $\nu=n_\sigma\text{mod}\,2$, the Z$_2$-topological invariant can equally be evaluated by computing the spin-Chern number $n_\sigma$ associated to the two-fold Harper equations through the twisted-periodic-boundary method~\cite{sheng:2006}.

\end{document}